\documentclass[a4paper,twocolumn,10pt,aps,pre]{revtex4-1}
\pdfoutput=1

\usepackage[intlimits]{amsmath}
\usepackage{amssymb}
\usepackage{amsfonts}
\usepackage{amsthm}
\usepackage{units}
\usepackage[caption=false]{subfig}
\usepackage{xspace}

\usepackage[english]{babel} 
\usepackage[utf8]{inputenc}
\usepackage{graphicx}
\usepackage{color}
\definecolor{mylinkcolor}{rgb}{0.5812,0.0665,0.0659} 
\definecolor{mycitecolor}{rgb}{0.075,0.31,0.0431} 
\definecolor{myurlcolor}{rgb}{0.0118,0.098,0.7412} 
\usepackage[pdftex,bookmarks,bookmarksopen,citecolor={mycitecolor},
linkcolor={mylinkcolor},urlcolor={myurlcolor},breaklinks=true,
hypertexnames=false,hyperindex=true,encap,colorlinks]{hyperref}

\def\scafacos{{\scshape Sca\-Fa\-CoS}}
\def\es{{\scshape ESP\-Res\-So}}
\def\memd{{MEMD}}
\def\pppm{{P3M}}

\newcommand{\plotwidth}{0.95\linewidth}

\usepackage{bm} 
\newcommand{\Vect}[1]{\ensuremath{\bm{#1}}} 

\newcommand{\ie}{\emph{i.e.\/}\xspace}

\newcommand{\icc}{ICC$\star$}

\begin{document}

\title{Computing Coulomb Interaction in Inhomogeneous Dielectric Media via a Local Electrostatics Lattice Algorithm}


\author{F.~Fahrenberger}
\email{Florian.Fahrenberger@icp.uni-stuttgart.de}
\author{C.~Holm}
\email{holm@icp.uni-stuttgart.de}
\affiliation{Institut f\"ur Computerphysik,\\ Universit\"at Stuttgart,\\ Allmandring 3, 70569 Stuttgart, Germany}

\begin{abstract}
  The local approach to computing electrostatic interactions proposed
  by A.~Maggs and adapted by J.~Rottler and
  I.~Pasichnyk for Molecular Dynamics simulations is
  extended to situations where the dielectric background medium is
  inhomogeneous. We furthermore correct a problem of the original
  algorithm related to the correct treatment of the global dipole
  moment, provide an error estimate for the accuracy of the algorithm,
  and suggest a different form of the treatment of the self-energy
  problem. Our implementation is highly scalable on many cores, and we
  have validated and compared its performance against theoretical
  predictions and simulation data obtained by other algorithmic
  approaches.
\end{abstract}


\maketitle

\section{Introduction}

A key component for dealing with larger scale soft matter systems is the coarse graining approach to model building. By reducing the degrees of freedom of the particle system, a significant speedup can be achieved allowing to reach the necessary length and time scales of soft matter systems. A very common approach to coarse graining in aqueous soft matter systems such as biological systems, polyelectrolyte solutions or colloidal suspensions \cite{roux99b,holm01a,holm04a} is to treat the solvent implicitly.  In most of these models water is not included as an explicit species, but is simply modelled as a homogeneous dielectric background that reduces the electrostatic interactions everywhere by the inverse of the relative dielectric constant, \ie 80 for water at room temperature.  While this assumption works well in homogenous solutions, it breaks down in the presence of spatial changes in the dielectric constant. This happens, for example, if walls are part of the aqueous system, \ie salty water in nano pores, glas walls in microfluidic channels, or the air -water interface of an open solution.

Many approaches have been presented in recent years to deal with dielectric enclosures, where interfaces between regions of different permittivity are treated via explicit image charges or boundary elements~\cite{linse86a,messina02f,boda04a,tyagi07a,tyagi08a,tyagi10a,xu13a,arnold13c}, via extended Poisson-Boltzmann solvers~\cite{honig93a,lu06a,altman09a}, or as an extension of a local Monte Carlo scheme~\cite{thompson08a}. While the influence of sharp dielectric contrasts in biophysical systems with a homogeneous salt concentration is not yet fully understood, its importance has been clearly demonstrated in such systems as ions near an air-water interface \cite{jungwirth06a}, where the inclusion of the correct polarizability plays a crucial role.
%
Another important step would be to take into account not only the effects of sharp dielectric contrasts, but allow for the bulk dielectric background to become inhomogeneous. Such effects can happen, for example in regions of water with different salt concentrations. We know that the dielectric constant of water depends on salt concentration, see for example ref \cite{hess06a,hess06b,bonthuis12a} and references therein. It becomes lower, as the concentration of salt is increased. Physically, this can occurs in many systems, \ie~polyelectrolyte solutions, suspensions of charged colloids, or the variation in polarizability of water close to strongly charged surfaces~\cite{bonthuis12a}. In addition, to our knowledge there is currently only one other recently presented algorithm~\cite{boda11a} that allows particles to cross a dielectric boundary and enter a region with different dielectric properties, and it requires more computational effort and is less versatile.

Smoothly varying dielectric properties have recently been the subject of algorithmic research and a few methods have been explored~\cite{buyukdagli13a,jadhao13a,levy12a,maggs12a,rottler11a,zwanikken13a}. In this article, an extension to the local Maxwell Equations Molecular Dynamics (\memd{}) algorithm is presented, the idea of which was first introduced by A.~Maggs~\cite{maggs02a,maggs04a} and later adapted by J.~Rottler and A.~Maggs~\cite{rottler04a}, and simultaneously by I.~Pasichnyk and B.~D\"unweg~\cite{pasichnyk04a} for molecular dynamics (MD) simulations. \memd{} is extended to deal with locally spatially varying dielectric properties for Coulomb interactions. We also point out general restrictions of the algorithm, and introduce an important correction to the handling of the global dipole term.

The \memd{} electrostatics approach has not been as widely adopted as many other well known electrostatics algorithms~\cite{arnold05a}, such as the various particle-mesh approaches~\cite{deserno98a} or the multipole method~\cite{greengard87a}. This is despite the fact that it comes with several benefits that have become of significant interest over the past few years. Since the Maxwell Equations for electrodynamics are intrinsically local and require no global information on the system, we gain two fundamental advantages:

First, unlike for all Ewald-based algorithms, the parallelization for such a local system of equations is trivial and communications need only be done at the boundaries of each domain. In addition, the scaling of the algorithm is only dependent on the lattice mesh size and therefore it scales linearly, $\mathcal{O}(N)$, for a fixed particle density. This is a very nice feature in a time where systems with several $10^7$ charges \cite{sonntag11a} are often simulated using massive parallel computers. Moreover this makes this algorithm also attractive to be ported on a graphical processing unit (GPU). Second, because of its locality the method easily allows changes of the dielectric properties within the system, which is introduced in this article.

The article is structured as follows. First, the algorithm (its initial and thermodynamic solution) is extended mathematically to locally varying permittivities. It is shown that the important features still hold and the statistical observables are reproduced correctly. Second, an estimate for the systematic error is presented and discussed. Third, the effect of handling the global dipole term in periodic boundary conditions is shown to be erroneous by a comparison to the classical Ewald method, paving immediately the way to constructing a correction term. Fourth, both the initial and the dynamic part of the extended algorithm are validated against analytical solutions and simulations. Fifth, the numerical performance of the algorithm is evaluated and advantages and limitations are discussed. Finally, we conclude and present a brief outlook on the future of the \memd{} algorithm.

\section{Extension of the algorithm}

Most of the proofs for the extended algorithm go along the lines of the original introduction by Maggs~\cite{maggs02a,maggs04a} Pasichnyk and D\"unweg~\cite{pasichnyk04a}. The general idea will be given, but the main mathematical steps can be retraced in the aforementioned publication. 

The algorithm consists of two parts: Calculating an initial solution of the Gauss law of electrodynamics on a lattice. The second part consists of applying and propagating all temporal changes to said solution within the system. The initial solution proposed by Pasichnyk for a constant permittivity only has to be adapted slightly to ensure the correct result for varying dielectric permittivities.

\begin{figure}
 \centering
 \includegraphics[width=0.7\linewidth]{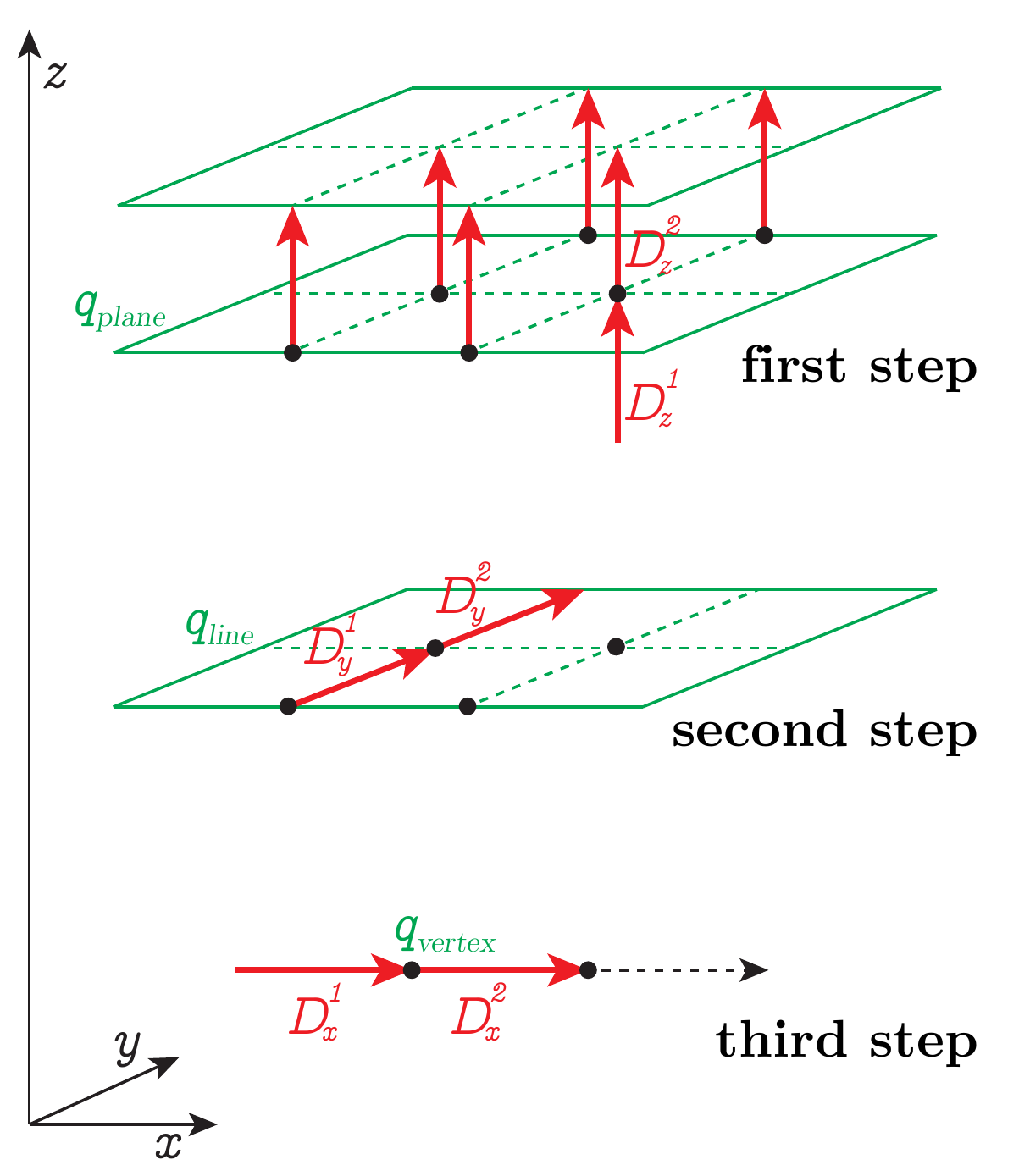}
 \caption{Recursive scheme for the initial solution of the E-field. The
  average charge in z-plane is scaled and added to each node, following
  $E_z^{(n+1)}=E_z^{(n)} + q_\text{plane}/(\varepsilon a^2)$. Then 
  the charge $q_\text{plane}$ is subtracted from each charge in the
  $z$-plane. Analogue with $y$-lines and the single nodes in
  $x$-direction.}
 \label{fig:initialscheme}
\end{figure}

A scheme to acquire an initial solution is shown in figure \ref{fig:initialscheme}. First, the charges on each plane are averaged, scaled by the lattice size and local permittivity, and added to the field on each node
\begin{equation}
E_z^{(n+1)}=E_z^{(n)} + \frac{q_\text{plane}}{\varepsilon_z^{x,y,n}a^2},
\end{equation}
and the charge $q_\text{plane}$ is subtracted from each vertex in the $z_n$-plane. The charges in $y$- and $x$-direction are updated accordingly on lines and vertices, following
\begin{align}
E_y^{(n+1)}&=E_y^{(n)} + \frac{q_\text{line}}{\varepsilon_y^{x,z,n}a^2}\\
E_x^{(n+1)}&=E_x^{(n)} + \frac{q_\text{vertex}}{\varepsilon_y^{y,z,n}a^2}.
\end{align}

Summation of the total charge in one cell is given by
\begin{equation}
q_\text{plane}+q_\text{line}+q_\text{vertex}
\end{equation}
and this yields the Gauss law directly, if the $(\nabla\cdot)$ operator is defined via finite differences $E^{(n+1)}-E^{(n)}$. An iterative procedure of energy minimization is equivalent to the second Maxwell equation $\nabla\times\Vect{E}=0$ and gives a correct initial solution. This method of numerical relaxation is not very efficient but has to be done only once.

Starting from this solution of Gauss' law, only updates of the electric field following a constraint have to be applied. Hereby, we can assume that the time scales of the propagation speed of the fields and the motion of the particles decouple. Then the propagation of the fields can be described by an artificial dynamics, in a Car-Parrinello (CPMD) manner~\cite{car85a}. Analog to Pasichnyk and D\"unweg, the most general constraint for the system is

\begin{equation}
\dot{\Vect{D}} + \Vect{j} - \nabla\times\dot{\Vect{\Theta}} = 0
\label{eq:constraint}
\end{equation}
with the electric displacement field $\Vect{D}=\varepsilon\Vect{E}$, the electric current density $\Vect{j}$, and an arbitrary vector field $\Vect{\Theta}$ as an additional degree of freedom. From this, the Lagrangian
\begin{equation}
\begin{split}
L =& \sum_i \frac{m_i}{2}\Vect{v}_i^2 - U \\
 & + \frac{f_\text{mass}}{2}\int \varepsilon(\Vect{r}) \dot{\Vect{\Theta}}^2d^3\Vect{r} -
           \frac{1}{2}\int\frac{\Vect{D}^2}{\varepsilon(\Vect{r})}d^3\Vect{r} \\
 & + \int\Vect{A}\left(\dot{\Vect{D}}-
           \nabla\times\dot{\Vect{\Theta}} + \Vect{j}\right) d^3\Vect{r}
\end{split}
\label{eq:lagrangian}
\end{equation}
is obtained, where the Lagrange multiplier $\Vect{A}$ is used to impose the kinematic constraint, $\Vect{r}$ is the position, $m_i$ and $\Vect{v}_i$ are the particle masses and velocities respectively, and $U$ is an additional potential. The prefactor $f_\text{mass}$ simply denotes the mass equivalent of the exchange particles, analog to electrons in CPMD, and later turns out to be related to the wave propagation speed as $1/c^2$.

The equations of motion for this Lagrangian $L(\Vect{r},\dot{\Vect{r}},\dot{\Vect{\Theta}},\Vect{D})$ can be calculated using variational calculus. The derivative in $\Vect{\Theta}$ and $\dot{\Vect{D}}$ is zero, and the motion of particles and fields is defined by
\begin{align}
\frac{d}{dt}\frac{\partial L}{\partial \dot{\Vect{r}}_i} - 
\frac{\partial L}{\partial \Vect{r}_i}  &\overset{!}{=} 0, \\
\frac{\partial\mathcal{L}}{\partial\dot{\Vect{\Theta}}} &\overset{!}{=} 0, \\
\frac{\partial\mathcal{L}}{\partial\Vect{D}} &\overset{!}{=} 0,
\end{align}
where $\mathcal{L}$ is the Lagrangian density, which by definition satisfies $L=\int\mathcal{L}d^3\Vect{r}$. Variation with respect to $\dot{\Vect{r}}_i$ results in

\begin{eqnarray}
\frac{\partial L}{\partial \dot{r}_i^\alpha} &=& m_i\dot{r}_i^\alpha + q_iA^\alpha (\Vect{r}_i) \notag \\
\frac{d}{dt}\frac{\partial L}{\partial \dot{r}_i^\alpha} &=& m_i\ddot{r}_i^\alpha + q_i\dot{A}^\alpha (\Vect{r}_i)
     + q_i\frac{\partial A^\alpha}{\partial r_i^\beta}\dot{r}_i^\beta \notag
\end{eqnarray}
where the second transformation is a time derivative. Variation with respect to $\Vect{r}_i$ yields

\begin{equation}
\frac{\partial L}{\partial r_i^\alpha} = -\frac{\partial U}{\partial r_i^\alpha}
     + q_i\dot{r}_i^\beta\frac{\partial A^\beta}{\partial r_i^\alpha}. \notag
\end{equation}
Combining these two results and introducing the vector field

\begin{equation}
\Vect{B} \mathrel{\mathop:}= \nabla\times\Vect{A}
\label{maggs:eq:B-vector}
\end{equation}
provides the equations of motion for the particle

\begin{eqnarray}
m_i\ddot{r}_i^\alpha &=& -\frac{\partial U}{\partial r_i^\alpha} - q_i\dot{A}^\alpha + q_i\dot{r}_i^\beta 
     \left( \frac{\partial A^\beta}{\partial r_i^\alpha} - \frac{\partial A^\alpha}{\partial r_i^\beta} \right)
     \notag \\
m_i\ddot{\Vect{r}}_i &=& -\frac{\partial U}{\partial\Vect{r}_i} - q_i\dot{\Vect{A}}
     + q_i \Vect{v}_i\times\Vect{B}
\label{maggs:eq:force}
\end{eqnarray}
as expected. The equations of motion for the electromagnetic fields can be found by varying the Lagrangian density $\mathcal{L}$. Variation in $\dot{\Vect{\Theta}}$ and in time gives

\begin{eqnarray}
\frac{\partial\mathcal{L}}{\partial\dot{\Vect{\Theta}}} &=& f_\text{mass}\varepsilon_0 \dot{\Vect{\Theta}}
     - \varepsilon_0 \nabla\times\Vect{A}
     = f_\text{mass}\varepsilon_0 \dot{\Vect{\Theta}} - \varepsilon_0 \Vect{B} \notag \\
\frac{d}{dt}\frac{\partial\mathcal{L}}{\partial\dot{\Vect{\Theta}}} &=& 
     f_\text{mass}\varepsilon_0 \ddot{\Vect{\Theta}} - \varepsilon_0\dot{\Vect{B}} = 0 \notag \\
f_\text{mass}\ddot{\Vect{\Theta}} &=& \dot{\Vect{B}}
\label{maggs:eq:propagate} \\
\frac{1}{c^2}\dot{\Vect{\Theta}} &=& \Vect{B},
\label{maggs:eq:wave}
\end{eqnarray}
where the natural initial condition $\dot{\Vect{\Theta}}(t=0) = 0$ is used in the last step, and $f_\text{mass}:=1/c^2$ for convenience. The next variation in $\Vect{D}$ gives

\begin{equation}
\dot{\Vect{A}} = -\frac{\Vect{D}}{\varepsilon},
\label{maggs:eq:DandA}
\end{equation}
which leads to the more commonly known expression for equation (\ref{maggs:eq:force}). With these two last results, (\ref{maggs:eq:wave}) and (\ref{maggs:eq:DandA}), two more Maxwell equations can be obtained by inserting into the constraint equation (\ref{eq:constraint}), namely Amp\`{e}re's and Faraday's law:

\begin{eqnarray}
\dot{\Vect{D}} &=& c^2\nabla\times\Vect{B} - \frac{\Vect{j}}{\varepsilon_0}
\label{maggs:eq:ampere}\\
\dot{\Vect{B}} &=& \nabla \times \dot{\Vect{A}} = -\nabla\times\Vect{D}.
\label{maggs:eq:faraday}
\end{eqnarray}

This means that simply applying the constraint (\ref{eq:constraint}) reproduces the complete electromagnetic formalism. It should be noted that the equations (\ref{maggs:eq:B-vector}) and (\ref{maggs:eq:DandA}) represent nothing more than the so-called temporal or Weyl gauge in electromagnetism, in which the scalar potential $\phi$ is identically zero, and which turns out to be the most appropriate gauge for our purposes.

Since the Lagrangian we introduced is constrained, it is not possible to easily construct a Hamiltonian from it, only via the Dirac theory of constrained systems. An elementary construction would be beneficial to simplify further proofs for the conservation of phase-space volume, energy and momentum. However, it is possible to construct a Lagrangian that is not constrained and produces the exact same equations of motion. The proofs and details will not be carried out, but the resulting Lagrangian is

\begin{equation}
\begin{split}
L =& \sum_i \frac{m_i}{2}\Vect{v}_i^2 - U + \frac{\varepsilon}{2}\int\dot{\Vect{A}}^2d^3\Vect{r} \\
 & -\frac{\varepsilon_0c^2}{2}\int (\nabla\times\Vect{A})^2d^3\Vect{r} + \int\Vect{A}\cdot\Vect{j}d^3\Vect{r}
\end{split}
\label{eq:unconstrained-lagrangian}
\end{equation}

The equations of motion for the particles and the fields can be derived from equation~\eqref{eq:unconstrained-lagrangian} and the Lagrangian density by the use of variational calculus. The resulting equations of motion for the particles and fields from the unconstrained Lagrangian are

\begin{align}
m_i\ddot{\Vect{r}}_i &= -\frac{\partial U}{\partial\Vect{r}_i} -q_i \Vect{E}
     + q_i \Vect{v}_i\times\Vect{B}
\label{eq:force}\\
\Vect{B} &= \frac{1}{c^2}\dot{\Vect{\Theta}}
\label{eq:theta}\\
\dot{\Vect{D}} &= c^2\: \nabla \times \Vect{B} - \Vect{j}
\label{eq:dfield}\\
\dot{\Vect{B}} &=  -\nabla\times\Vect{D}.
\label{eq:bfield}
\end{align}

In our implementation, the magnetic part of the Lorentz force, $\Vect{v}\times\Vect{B}$ from equation \eqref{eq:force}, is omitted. This increases the speed significantly, but makes it impossible to construct an unconstrained Lagrangian from eq.~\eqref{eq:unconstrained-lagrangian}, and therefore the Hamiltonian nature of the algorithm does not hold. Momentum conservation is violated by the amount of momentum that the virtual photons carry. This is a negligible percentage and perfect momentum conservation is not important in most simulated systems since many contain a thermostat. Energy conservation, however, holds, as can be shown with a pseudo Liouville theorem for the Lagrangian, along the lines of~\cite{pasichnyk04a}. All additional terms that show up due to a spatially dependent permittivity simply cancel out.

Like in the algorithm for constant background permittivity, the thermodynamic observables are perfectly reproduced, since they are not dependent on the speed of light nor the magnetic field component. In contrast to the original version, the partition function in this extended algorithm contains an extra term for the varying permittivity. The particle momenta and the vector field $\Vect{A}$ can still be integrated out in a straightforward way. If we split up the integration of the electric field in a longitudinal and a transversal component, we end up with

\begin{align}
\mathcal{Z} =& \int \prod_{i=1}^{N}d\Vect{r}_i\prod_{\Vect{r}}
\mathcal{D}\Vect{D}^{\|}(\Vect{r})\mathcal{D}\Vect{D}^{\bot}(\Vect{r})\:
\delta\left(\nabla\cdot\Vect{D} - \rho(\Vect{r})\right) \notag \\
 & \quad \cdot \exp\left(-\frac{\beta}{2}\int d\Vect{r}\frac{\Vect{D}^{\|}(\Vect{r})^2}{\varepsilon(\Vect{r})}\right) \notag \\
 & \quad \cdot \exp\left(-\frac{\beta}{2}\int
  d\Vect{r}\frac{\Vect{D}^{\bot}(\Vect{r})^2}{\varepsilon(\Vect{r})}\right)
\label{eq:partitionfunction}.
\end{align}

The integration over the transversal component also only contributes a factor, and the longitudinal component cancels with the delta function. This contribution of the transversal component is constant for a static dielectric background, but can vary if the dielectric interfaces are mobile. This gives rise to thermal Casimir/Lifshitz interactions as discussed by Pasquali and Maggs~\cite{pasquali08a,pasquali08b,pasquali09a}, but the effect will not be discussed further in this article since we focus on moving charges in static dielectric backgrounds. The only degrees of freedom now left are the particle coordinates, which finally leads to

\begin{align}
\mathcal{Z} &=& \int \prod_{i=1}^{N}d\Vect{r}_i \exp\left(-\frac{\beta}{2}\int
  d\Vect{r}\frac{\Vect{D}(\Vect{r})^2}{\varepsilon(\Vect{r})}\right).
\label{eq:finalpartitionfunction}
\end{align}
This is what is expected from the static case of electromagnetic interactions.

\begin{figure}
\subfloat[discretization][Discretization]{
 \includegraphics[width=0.5\linewidth]{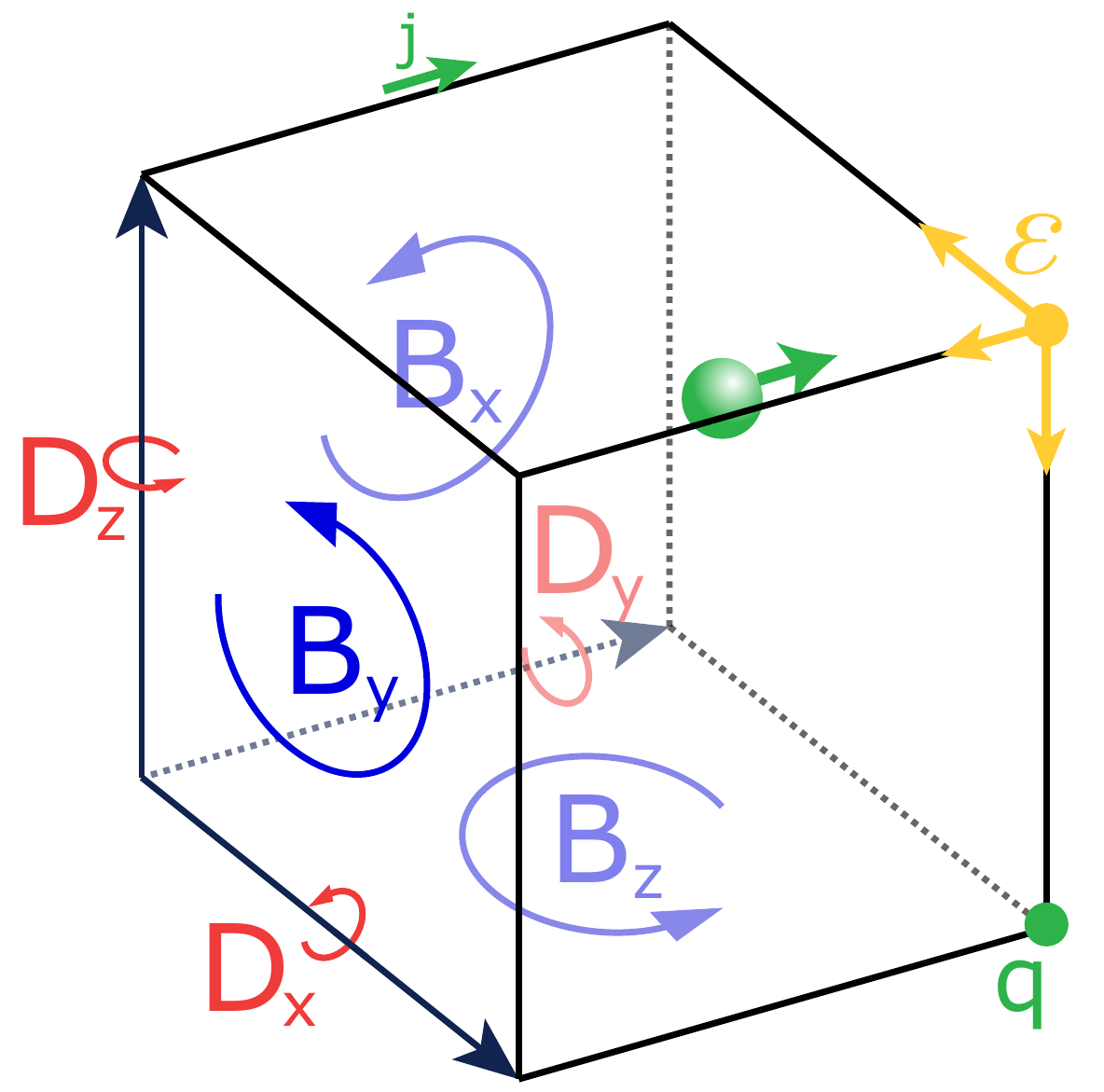}
 \label{fig:discretization}
}
\subfloat[discretization][$\varepsilon$ interpolation]{
 \includegraphics[width=0.5\linewidth]{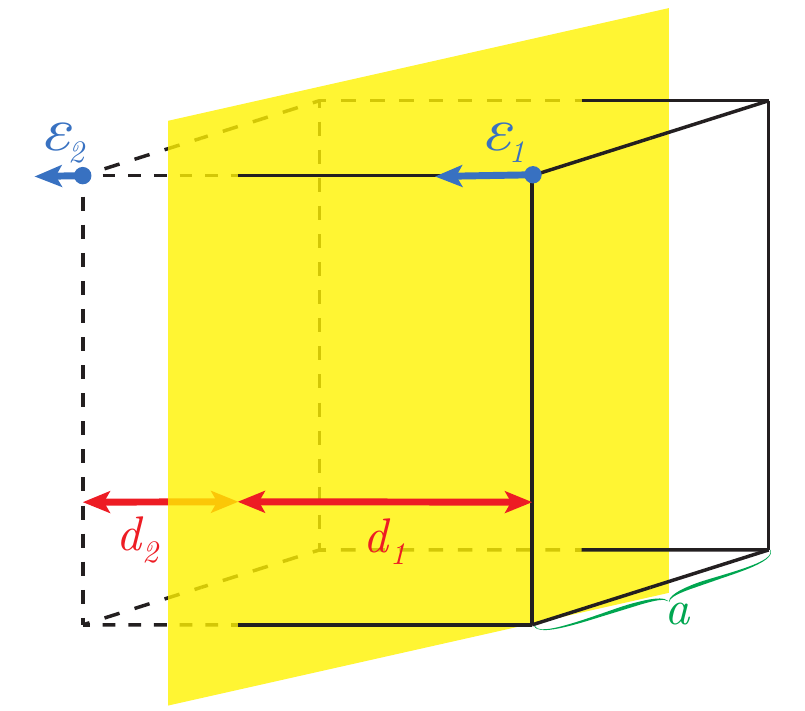}
 \label{fig:epsilon-interpolation}
}
 \caption{(a) Discretization of the currents, fields, and permittivities onto a lattice cell. (b) Interpolation of dielectric permittivity values on the lattice. $\varepsilon(\Vect{r})$ has a position and a direction (blue arrow). The values for $\varepsilon_1$ and $\varepsilon_2$ are determined and the value on the connecting link is set to the average value. If the gradient is too large, the value is determined by forming the harmonic average.}
\label{fig:discretizations}
\end{figure}

The lattice discretization in space is done in a way analog to the original implementation (see figure~\ref{fig:discretization}), featuring the same finite differences representations for the gradient $(\nabla\cdot)$ and the curl $(\nabla\times)$ operators. The local permittivity values $\varepsilon$ can assume tensorial form, equivalent to a differential 2-form. In our implementation, we reduce the tensor to its diagonal entries (differential 1-form), which merely represents an optically isotropic medium. The local permittivity therefore has a value and a direction, and they are placed on the links of the interpolating grid. The electric displacement field values $\Vect{D}=\varepsilon\Vect{E}$ are still stored on the links, although they are represented by a rotation around these links.

To map given permittivity values, set by an interface or function, to the lattice, the finite difference between adjacent grid points is employed. If the difference is significantly bigger than the values, the link is marked as an \emph{interface link}. The values for these interface links are then calculated by taking the harmonic average
\begin{equation}
\varepsilon_\text{link} = \varepsilon_1\cdot\frac{d_2}{a} + \varepsilon_2\cdot\frac{d_1}{a},
\label{eq:dielectric-interpolation}
\end{equation}
where $\varepsilon_1$ and $\varepsilon_2$ are the permittivity values on the adjacent lattice sites on each side of the interface respectively, $d_1$ and $d_2$ are the distances of the according lattice site along the link to the interface, and $a$ is the lattice spacing, as depicted in Fig.~\ref{fig:epsilon-interpolation}.

\section{Self energy interaction}

Even in the continuum, the solution of the Maxwell equations for point charges is singular at the position of the particle. The point charge carries along with it the electrostatic energy

\begin{equation}
\frac12 \int_{|\Vect{r}-\Vect{r}_i(t)|\leq R} \frac{\Vect{D}(\Vect{r},t)^2}{\varepsilon} d^3\Vect{r}
   \propto \int_0^Rr^2(r^{-2})^2 dr = \int_0^Rr^{-2}dr \notag
\end{equation}
which is a diverging integral. This would mean that the particle has infinite mass and can not respond to forces. With a lattice spacing, a ``cut-off'' is introduced for this self-interaction, but still the particle is driven to the center of the cell by the field created from its own (interpolated) charge. It is, from an energy point of view, most favorable for the particle to distribute its charge evenly on all surrounding lattice points, since it then produces the smallest possible curl $(\nabla\times\Vect{D})$ in the cell.

This spurious self-influence is in the original algorithm corrected by the use of Yukawa-potentials. However, if the permittivity of the system changes within the cell, every potential based correction scheme fails. However, this problem can be solved both with a lattice Greens function, or a direct subtraction scheme.

In our implementation, the permittivity, as it is a differential 1-form, is placed on the lattice links. Therefore it remains constant on the link, allowing us to set up a Green's function of the form

\begin{align}
\Delta_{\Vect{r}'}G(\Vect{r}-\Vect{r}') =
-\frac{1}{a^2}\delta_{\Vect{r}, \Vect{r}'}
\label{eq:greens}
\end{align}
if $\Vect{r}$ and $\Vect{r}'$ are placed on adjacent lattice sites. For a point charge, the electrostatic potential can then be found using a convolution with the Green's function

\begin{align}
\phi(\Vect{r}) = \frac{q}{a\varepsilon G(\Vect{r})},
\end{align}

where we assume that $\varepsilon$ remains constant within the cell. This Green's function for a point charge on a cubic lattice can now be solved by a Fourier transform and is limited to the first Brioullin zone. For an infinitely large lattice, the back transform yields the integral

\begin{align}
G(\Vect{r}) = \left(\frac{a}{2\pi}\right)^3 \int_{\Vect{k}\in
  \text{BZ}} \frac{e^{i\Vect{k}\Vect{r}}}{\varepsilon(\Vect{k})} d^3k.
\end{align}

If the Laplace operator on the left-hand side of equation \eqref{eq:greens} is used to construct a finite-differences operator and applied to each of the interpolated charges on the lattice, we end up with a solution of the self-energy influence that can be added up. This is a well-known scheme for lattices of constant dielectric permittivity and applies in a straight forward way here if $\varepsilon$ does not change within the cell of each charge~\cite{joyce94a}. While this is not suitable for all cases, it can be applied to many and is fairly fast, since the solution for the given integral only needs to be calculated once at the beginning of the simulation and it can be done analytically~\cite{glasser00a}.

Another approach to allow for dielectric variations on very small scales (within one lattice site) is to use a direct subtraction scheme. With the assumption that the gradient of the permittivity, $\nabla\varepsilon(\Vect{r})$, is constant on each lattice link, the influence of the interpolated charges can be directly calculated and subtracted. This also requires the charge interpolation scheme to be of linear order and, after some algebra, results in

\begin{align}
\Vect{E} = \sum_{d=x,y,z} \sum_{\substack{i,j=0\\ i\neq j}}^{1} 
\varepsilon(r_{j,d}-r_{i,d})
\frac{q\cdot{}a\cdot{}(r_{i}r_{i,d} + r_{j}r_{j,d})}{r_{i}r_{j}r_{i,d}r_{j,d}}
\label{eq:direct-subtraction}
\end{align}
where $q$ denotes the charge of the particle, $a$ the lattice spacing, $r_{i,d} = |\Vect{r}-\Vect{r}_{i,d}|$ the position of the charge relative to the vertex~$(i,d)$, and $r_{i} = |\Vect{r}\cdot(\Vect{r}_{j,d}-\Vect{r}_{i,d})|$ the position of the charge folded onto dimension~$d$ relative to the vertex~$(i,d)$, and $\varepsilon(r_{j,d}-r_{i,d})$ the permittivity on the lattice link between vertices~$(i,d)$ and~$(j,d)$.

Both correction schemes as well as the use of Yukawa potentials are included in our implementation within \es{}~\cite{arnold13a,limbach06a}, an Extensible Software Package for Research on Soft Matter. For constant dielectric background, Yukawa potential correction is used as the most precise option. For dielectric interfaces that can not be approached and penetrated by charges, the Greens function correction is suitable. The direct scheme is the most general, and it is used by default for spatially varying dielectric systems.

\section{Error estimates}
\label{sec:error}

An error estimate for the local electrostatics algorithm has been presented in~\cite{rottler07a}, but it features a more complex interpolation scheme and can not be applied here. Thus, we want to analyze and check the error contributions of our implementation. Other than the physical error from omitting the magnetic part of the Lorenz force, the \memd{} algorithm carries a numerical error which consists of two parts.

The first contribution stems from the linear interpolation scheme of the electric current onto the lattice. It can be reduced by introducing a splitting of the far field and near field part of the Coulombic interaction and calculating the near field (e.g. the 27 surrounding next neighbor lattice cubes) with a direct pairwise potential calculation. This splitting, however, comes with a considerable overload in computational effort, since the near field correction needs to be subtracted from the \memd{} algorithm via a lattice Greens function or a similar construct. It is also not possible for spatially varying dielectric properties, since the Coulomb potential based near field approach breaks down for a non-constant dielectric background. This is why, for the extended algorithm, we stick with a linear interpolation scheme with a short-range cutoff of only one lattice cell.

The second contribution is of algorithmic origin and relates to the artificially small speed of light. Since the propagation speed of the magnetic fields is finite, the system does not feature true electrostatics but retarded solutions of the Maxwell equation. This error is also indirectly related to the lattice spacing, since a coarser lattice allows for the magnetic fields to be propagated over greater distances in one time step.

The two errors are connected via the Courant stability criterion:

\begin{align}
c & \ll  \frac{a}{dt}. \label{eq:stability-criterion}
\end{align}

\begin{figure}
 \centering
 \includegraphics[width=\plotwidth]{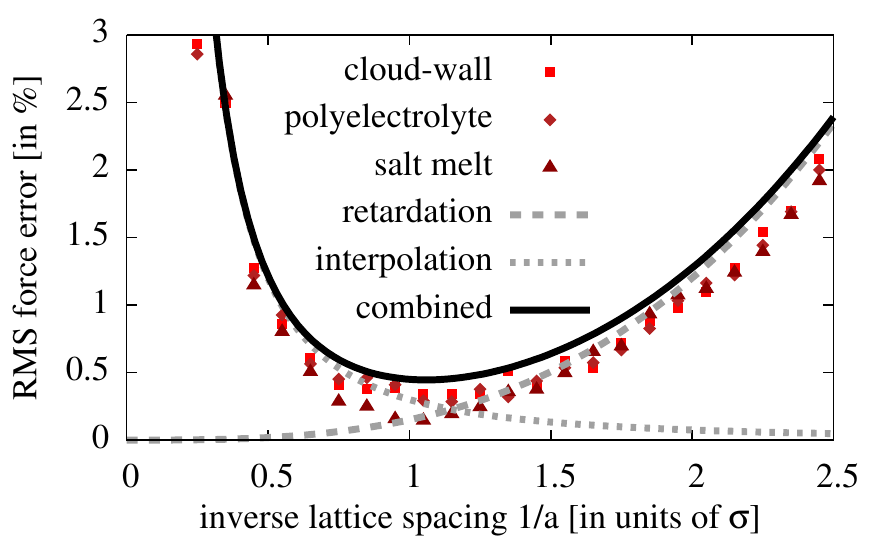}
 \caption{Error estimate for the \memd{} algorithm: The interpolation and the finite
 speed of light create numerical errors. Both depend on the lattice
 spacing. For affirmation, the errors from three simulations of different systems (see fig.~\ref{fig:error-systems}) are included, compared to high precision P3M force calculations.}
 \label{fig:theoretical-error}
\end{figure}

With a lattice spacing $a$, the error introduced by a linear interpolation scheme for geometric reasons scales with $1/a^3$, whereas the algorithmic error scales with $a^2$ (from equations \eqref{eq:dfield} and \eqref{eq:stability-criterion}). The resulting overall numerical error is shown in figure~\ref{fig:theoretical-error} for a random distribution of charges. In addition, three simulations have been performed with the \memd{} algorithm and compared in force to reference values from a \pppm{} implementation tuned to high accuracy.
 
\begin{figure}
\centering
\subfloat[error systems][polyelectrolyte]{
  \includegraphics[width=0.45\linewidth]{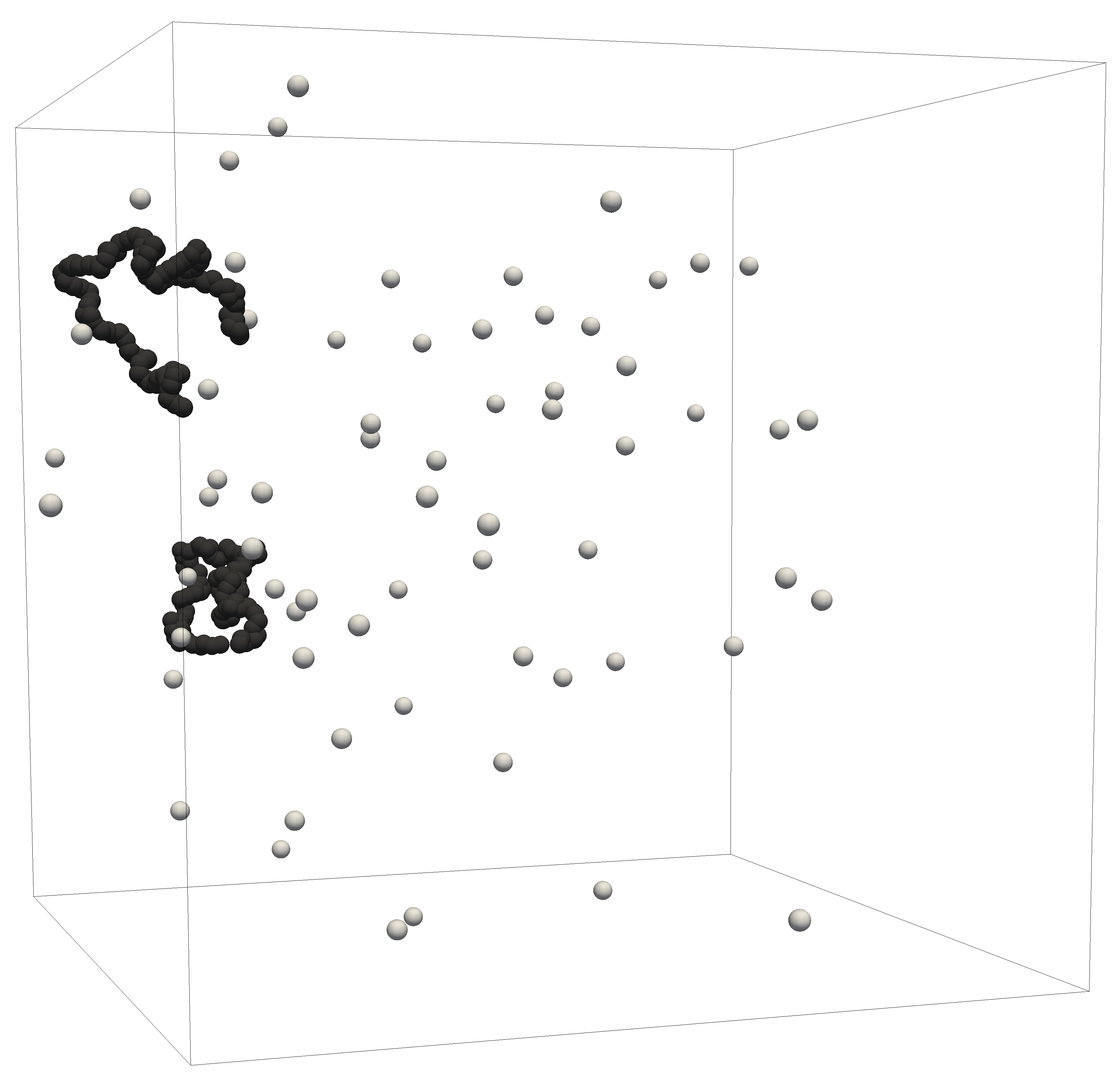}
  \label{fig:polyelectrolyte}
}\\
\subfloat[error systems][salt melt]{
  \includegraphics[width=0.45\linewidth]{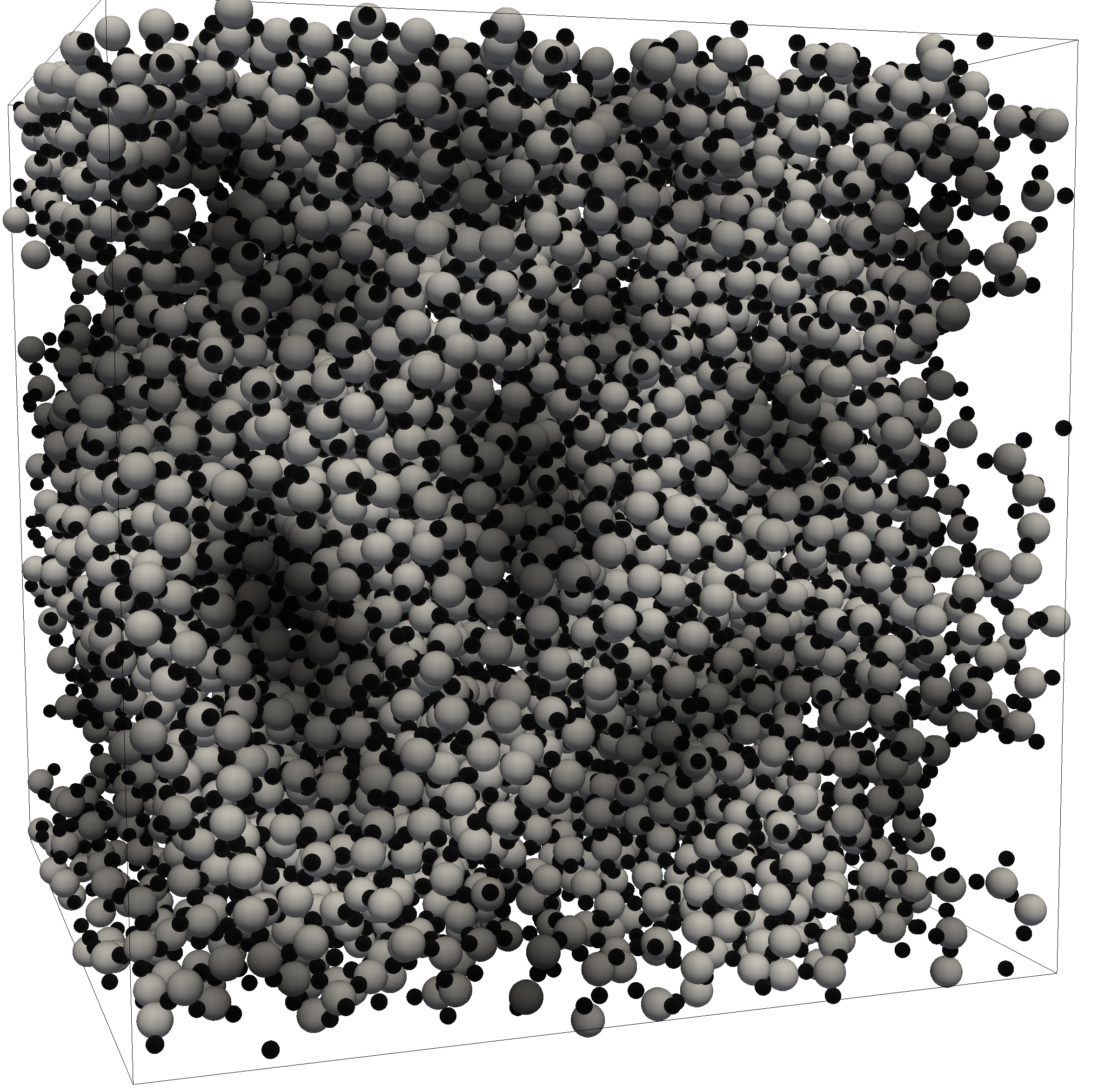}
  \label{fig:salt-melt}
}
\subfloat[error systems][cloud-wall]{
  \includegraphics[width=0.45\linewidth]{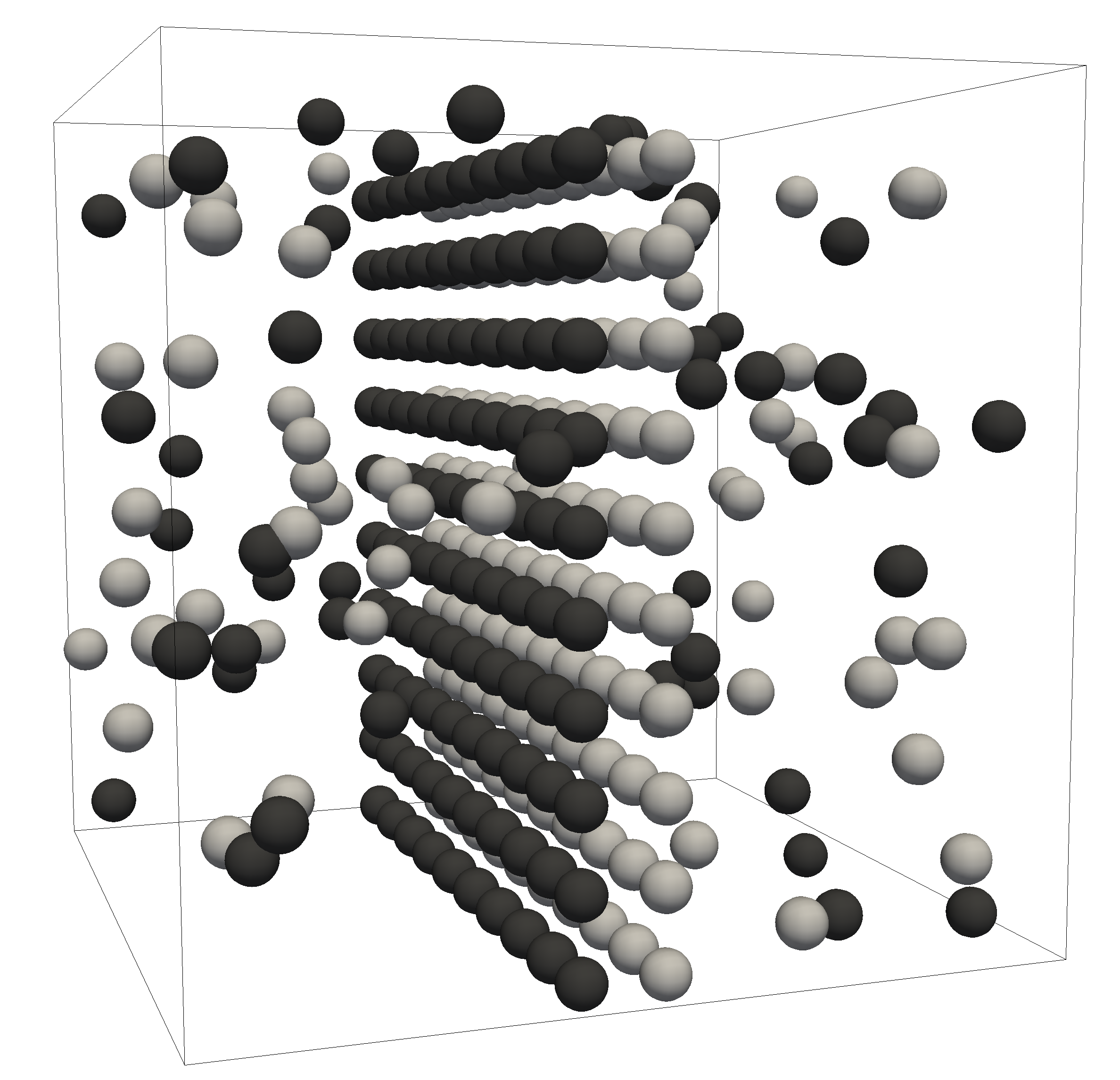}
  \label{fig:cloud-wall}    
}
\caption{Three example systems to determine the numerical error of the algorithm. Two polyelectrolytes in aqueous solution (a), a melting silica crystal (b), and an artificial system (c) with two oppositely charged walls and a surrounding cloud of charges.}
\label{fig:error-systems}
\end{figure}

The three systems were chosen to represent three different setups that contain different types of error sources. The system of a polyelectrolyte in salt water solution in figure~\ref{fig:polyelectrolyte} features a very dilute system with a high charge concentration around the polymer. Oppositely, the silica melt in figure~\ref{fig:salt-melt} is very dense and includes ions of different valency. The third example is an artificial setup of two infinite charged walls and a surrounding cloud of randomly placed charges. This system features a strong dipole moment and a significant long-range contribution throughout the simulation box.

From figure~\ref{fig:theoretical-error}, two things can be deduced: First, the numerical error has a predictable minimum, since the two error sources have a clear mathematical relation. Second, the relative RMS force error for a typical system does not go below $10^{-3}$. This is sufficient for most MD simulations, but it should be considered before using this algorithm. The error can be reduced further by introducing an effective short range cutoff, as mentioned above.

It should also be noted that the error increases at very small distances between two charges due to the linear interpolation. For inhomogenous systems with very dense areas not only the error increases but the algorithm slows down significantly, having to propagate all fields into empty regions.

\section{Periodicity}
\label{sec:periodicity}

In MD simulations, the box geometry is often set to be periodic in all dimensions, to avoid boundary effects. This type of boundary condition is introduced very naturally in the \memd{} algorithm. Because of its locality, the boundaries and according field propagations can be directly stitched together, creating infinitely many periodic replicas.

In fully periodic charged systems using Ewald-based algorithms, like the particle-particle particle-mesh method \pppm{}, the boundary conditions at infinity have a considerable effect on the solution~\cite{deleeuw80a}. For physical simulations, one normally assumes metallic boundaries at infinity to fix the electric field and potential to zero. This assumption cancels the system's overall charge and its dipole moment for force calculations, since the potential and the electric fields at the metallic boundary at infinity are forced to zero. Physically, this allows the algorithm to compensate for a non-neutrality of the system, and it allows the dipole moment of the system to perform an unrestricted random walk. This is intentional, because in realistic systems, the dipole moments of all periodic copies would not be exactly the same but perform individual random walks to cancel out on average after spatial integration~\cite{neumann83d}.

For a standard Ewald method, the correction of the dipole term to the electric field at $\varepsilon(\infty)=1$ has been calculated in~\cite{deleeuw80a} and~\cite{caillol94a} as
\begin{equation}
\mathcal{H}_\Delta = - \frac{\rho(\varepsilon'-1)}{2N(2\varepsilon'+1)}\sum_{1\le i<j\le N}\Vect{\mu}(i)\cdot\Vect{\mu}(j),
\end{equation}
where $\rho$ is the charge distribution, $N$ is the number of particles, $\Vect{\mu}$ are magnetic dipoles in the system, and $\varepsilon'$ is the permittivity at infinite distance. If this is transferred to electrostatic monopoles, it will create an energetic influence $\Phi_\Delta$ on each particle of

\begin{equation}
\Phi_\Delta = \frac{\Vect{P}^2}{2(2\varepsilon_b + 1)\varepsilon_0 V},
\end{equation}

where $V$ is the volume of the simulation box, $\varepsilon_b$ is the boundary permittivity at infinite distance, and $\Vect{P}$ is the total dipole moment of the unfolded coordinates of the charges. Similar corrections have been introduced in~\cite{levrel08a,rottler04b} with additional moves to the original Monte Carlo algorithm, but were not transferred to the MD implementation.

In Ewald-based methods, the monopole and dipole terms can directly be set to zero in Fourier space, providing a simple and exact way to set metallic boundary conditions at infinity. This is not possible within the \memd{} algorithm since it solves the Maxwell equations locally in real space. In addition, the contribution in \memd{} will not only include the dipole moment contribution of the folded particle coordinates within the box geometry but also that of the unfolded coordinates, since the phase space history of the system is stored in the magnetic field.

\begin{figure}
\subfloat[dipole moment][Dipole moment fluctuations]{
 \includegraphics[width=\plotwidth]{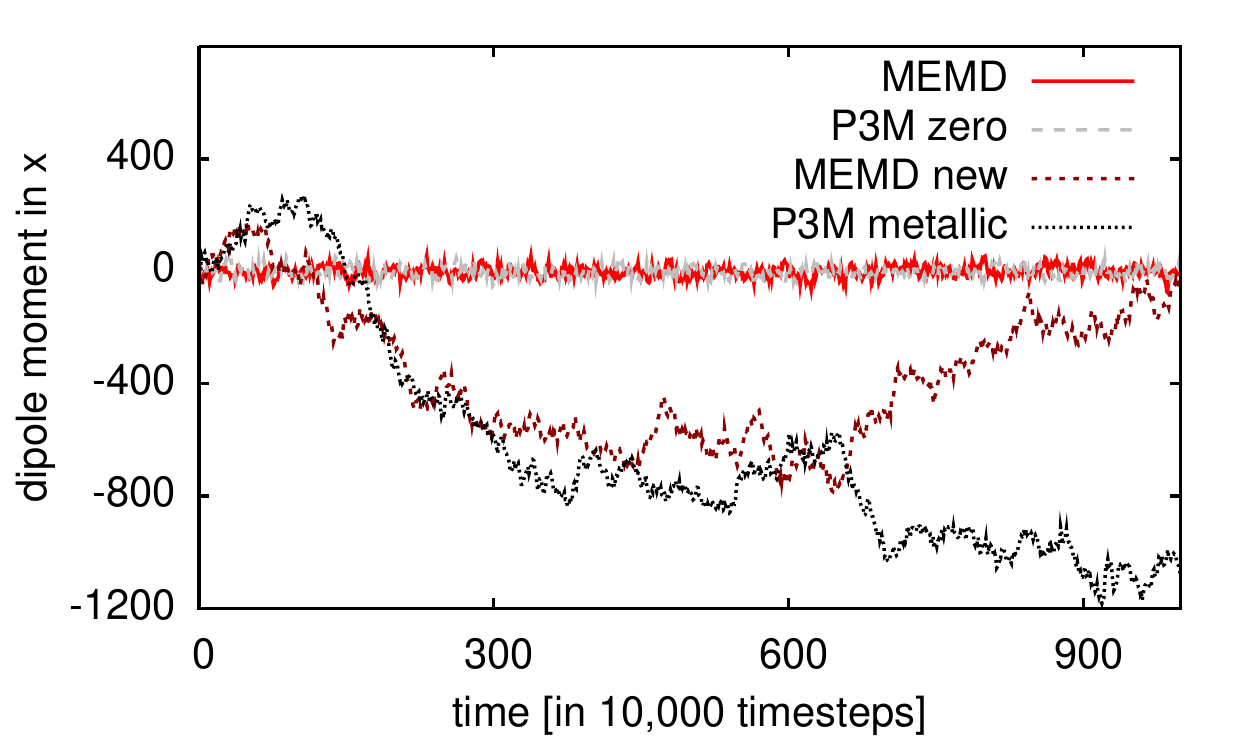}
 \label{fig:dipole-moments-fluctuation}
}\\
\subfloat[dipole moment][Force dependence on the dipole moment]{
 \includegraphics[width=\plotwidth]{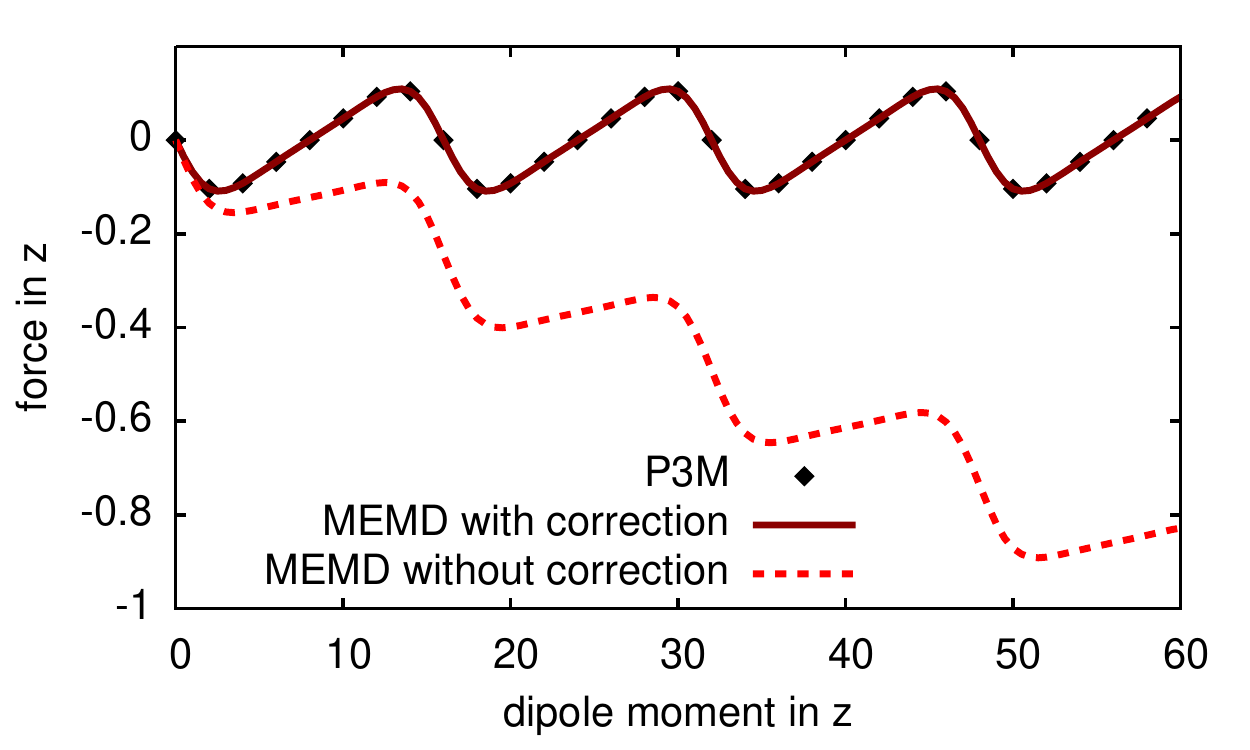}
 \label{fig:dipole-moment-force}
}
\caption{(a) The dipole moments of a simple electrolyte system are compared for the \memd{} and a \pppm{} algorithm. The original \memd{} keeps the dipole moment at zero, which corresponds to \pppm{} boundary conditions of $\varepsilon = 0$ at infinity. The corrected version shows the same behavior as \pppm{} for metallic boundary conditions. (b) Two particles are dragged apart over several lengths of the simulation box via applying an external field. As the dipole moment of the system increases, these two boundary conditions also diverge quantitatively in the force calculation.}
\label{fig:dipole-moments}
\end{figure}

For clarification, two simulations were set up, and the results can be seen in figure~\ref{fig:dipole-moments}. First, an electrolyte was simulated over a long time and the total dipole moment of the system was recorded for the original and the corrected \memd{} implementation, as well as \pppm{} with metallic and $\varepsilon=0$ boundaries (figure~\ref{fig:dipole-moments-fluctuation}). For the corrected local algorithm and \pppm{} with metallic boundaries, a random walk of the dipole moment can be observed, as expected for an unrestricted system. The uncorrected algorithm shows the same behavior as \pppm{} with $\varepsilon=0$ boundaries, forcing the dipole moment to remain around zero in a harmonic potential. Second, two charges of opposite sign are placed in a system and slowly pulled apart by an external field (figure~\ref{fig:dipole-moment-force}). For the uncorrected algorithm, the oscillating force on one particle induced by the periodic images of the second particle is overshadowed by the influence of the resetting dipole force for the linearly increasing dipole moment of the system.

As can be seen in figure~\ref{fig:dipole-moments-fluctuation}, this true 3D periodic behavior of \memd{} relates to boundary conditions of $\varepsilon = 0$ at infinity in the \pppm{} equivalent. This is an unwanted effect, and in a local real space method, there is no mathematical trick to apply metallic boundary conditions as in Ewald methods. The solution in our implementation is to calculate the system's dipole moment directly from the unfolded particle coordinates and subtract its influence from the force. This works reasonably well as long as there is no external driving force on the dipole moment. Systems with an additional external field or other means of introducing a net electric current will result in a drift in the dipole moment to a point where the dipole correction outweighs the actual force on the particles, and the algorithm breaks down. In these cases, the system has to be re-initialized quite often, and the badly scaling numerical relaxation leads to slow performance. Therefore, systems with external driving forces on charged particles should be avoided for \memd{}.

\section{Validation}

Two changes to the algorithm have to be validated in simulations: The
adapted scheme for our initial solution and the extended dynamic
algorithm for varying permittivity. For the initial scheme, we check
two setups against their analytical solution. Since there is no
analytical solution for the dynamic behavior of such a system, we
will investigate the distribution of particles in statistical equilibrium
for a system that has been studied with Monte Carlo simulations~\cite{messina02f} and can also be simulated with an algorithm for
dielectric interfaces~\cite{tyagi10a,kesselheim10a}.

\subsection{Initial solution}

To validate the initial solution for spatially varying dielectric media, the field is compared to the analytical solution in a system that consists of a single charge in the box center and a dielectric background that is $\varepsilon=1$ at the charge position and linearly rises with distance ($\varepsilon(r)=|r|$). This can be solved analytically using direct spherical integration.

\begin{figure}
\subfloat[linear dielectric][Potential]{
  \includegraphics[width=0.9\linewidth]{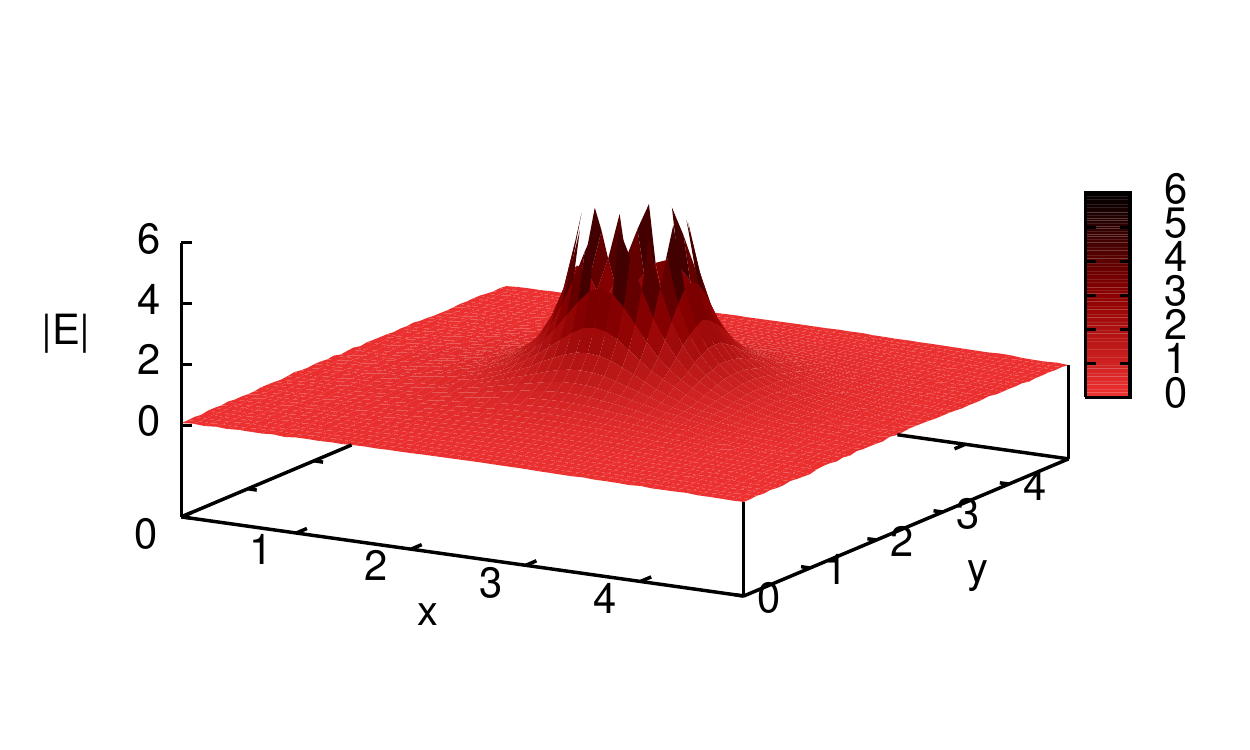}
  \label{fig:linear-dielectric-data}
}\\
\subfloat[linear dielectric][Relative errors]{
  \includegraphics[width=0.9\linewidth]{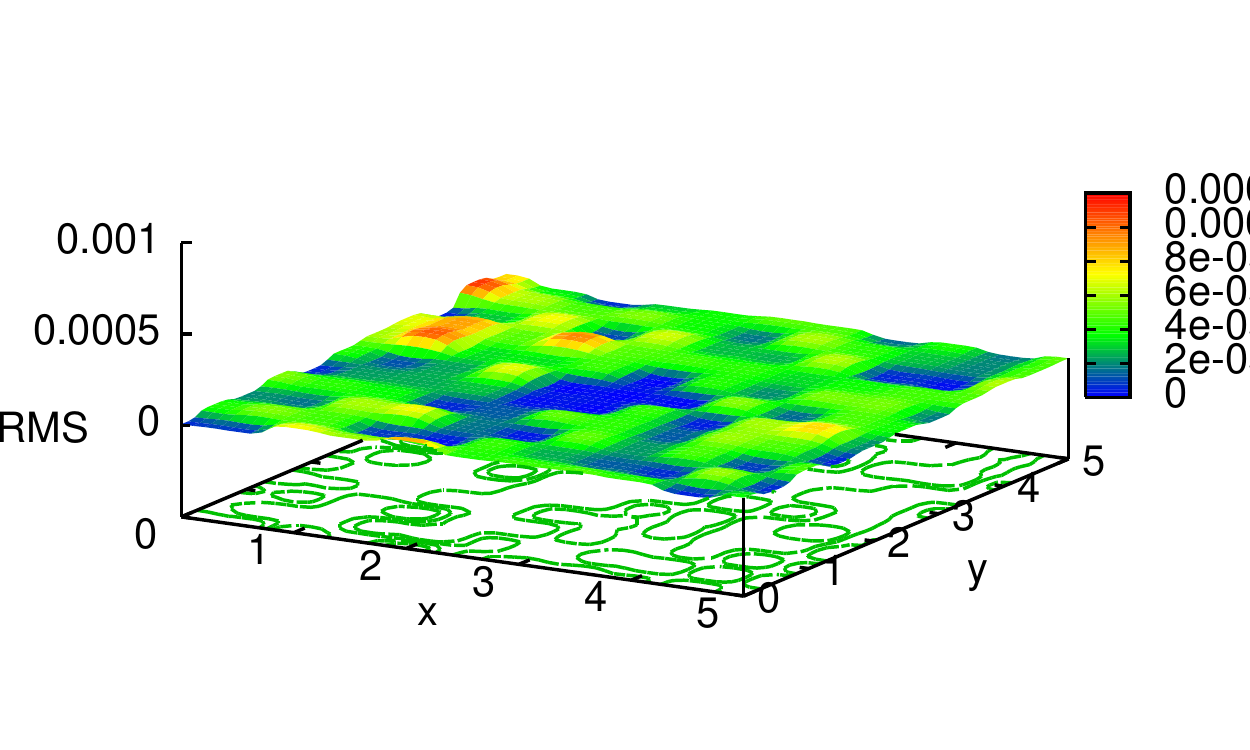}
  \label{fig:linear-dielectric-error}
}
\caption{(a) A charge is placed in the center, with the dielectric constant linearly rising with distance from the center. This graph shows the absolute value $\sqrt{\Vect{E}^2}$ of the resulting field. As theoretically predicted, the system follows a $1/r^3$ behavior. (b) The relative RMS error of the electric field. The graph shows that the error stays well below the $10^{-3}$ limit, and this relative error is smallest in the cell center.}
\end{figure}

The result can be inspected in figure~\ref{fig:linear-dielectric-data} and matches the analytic prediction. The relative RMS error of the absolute field value stays well below the algorithmic precision limit of $10^{-3}$, which is not surprising since the error contribution of the retarded solutions due to the dynamic algorithm is not present yet. While the absolute error shows no spatial preference, the relative error shown in figure~\ref{fig:linear-dielectric-error} is very small in the cell center and increases near the boundaries. This behavior shows that the absolute error is independent of the local field strength and only influenced by the lattice spacing.

\subsection{Thermodynamic behavior}

\begin{figure}
\centering
\subfloat[walls][System setup]{
 \includegraphics[width=0.5\linewidth]{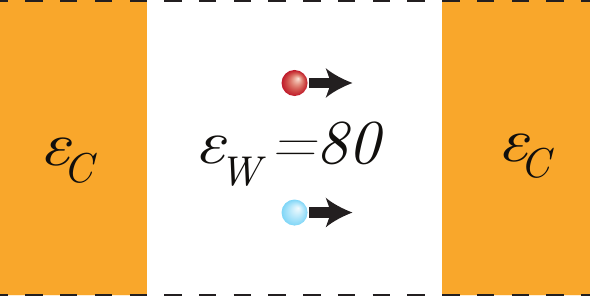}
 \label{fig:dielectric-wall-schematic}
}\\
\subfloat[walls][Force on the particle]{
 \includegraphics[width=\plotwidth]{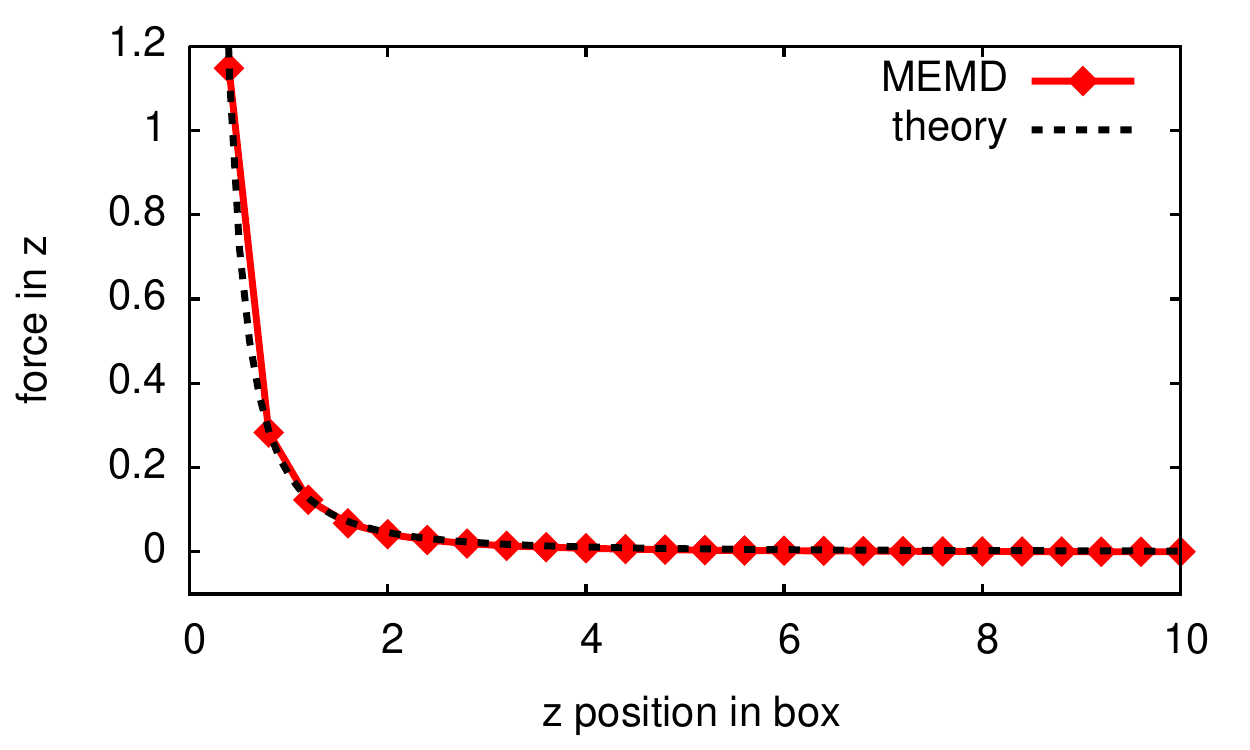}
 \label{fig:dielectric-wall-data}
}
\caption{(a) Two particles are simultaneously pulled through the simulation box ($\varepsilon_W=80$) with two infinite dielectric walls ($\varepsilon_C=2$). (b) The results are compared to the analytical prediction. The box center is located at $z=10$, and one wall at $z=0$ on the left of the graph.}
\label{fig:dielectric-wall}
\end{figure}

In order to validate the thermodynamic behavior of the extended algorithm, we place a charged particle between two dielectric walls and manually drag it from the center to one of the walls. In this artificial case, the force on the particle can be theoretically predicted in every position and compared to the simulation. The results can be seen in figure~\ref{fig:dielectric-wall} and match the analytical prediction up to algorithmic precision.

A less conceptional and more physical simulation is performed with a fixed colloid (charge $Q_c=60$, radius $R_c$, hard sphere boundary) in a solution of counterions and salt (concentration $c=\unit[50]{mmol/l}$). The radial distribution of the ions around the colloid is measured and compared against the results of a Monte Carlo simulation from~\cite{messina02f} and an MD simulation using the \icc{} algorithm for sharp dielectric boundaries~\cite{tyagi10a,kesselheim10a}. The data matches very well, but shows slight deviation at short distances from the boundary. This is due to the linear lattice interpolation of charges which leads to inaccuracy where charges are very close to each other, at the order of one lattice spacing.

\begin{figure}
\centering
\subfloat[colloid RDF][System setup]{
 \includegraphics[width=0.6\linewidth]{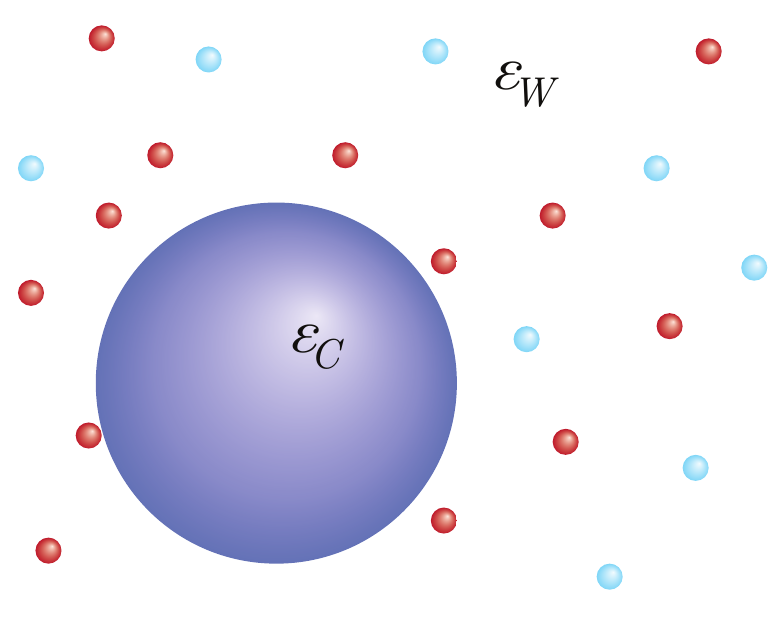}
 \label{fig:dynamic-sphere-scheme}
}\\
\subfloat[colloid RDF][Counterion radial distribution]{
 \includegraphics[width=\plotwidth]{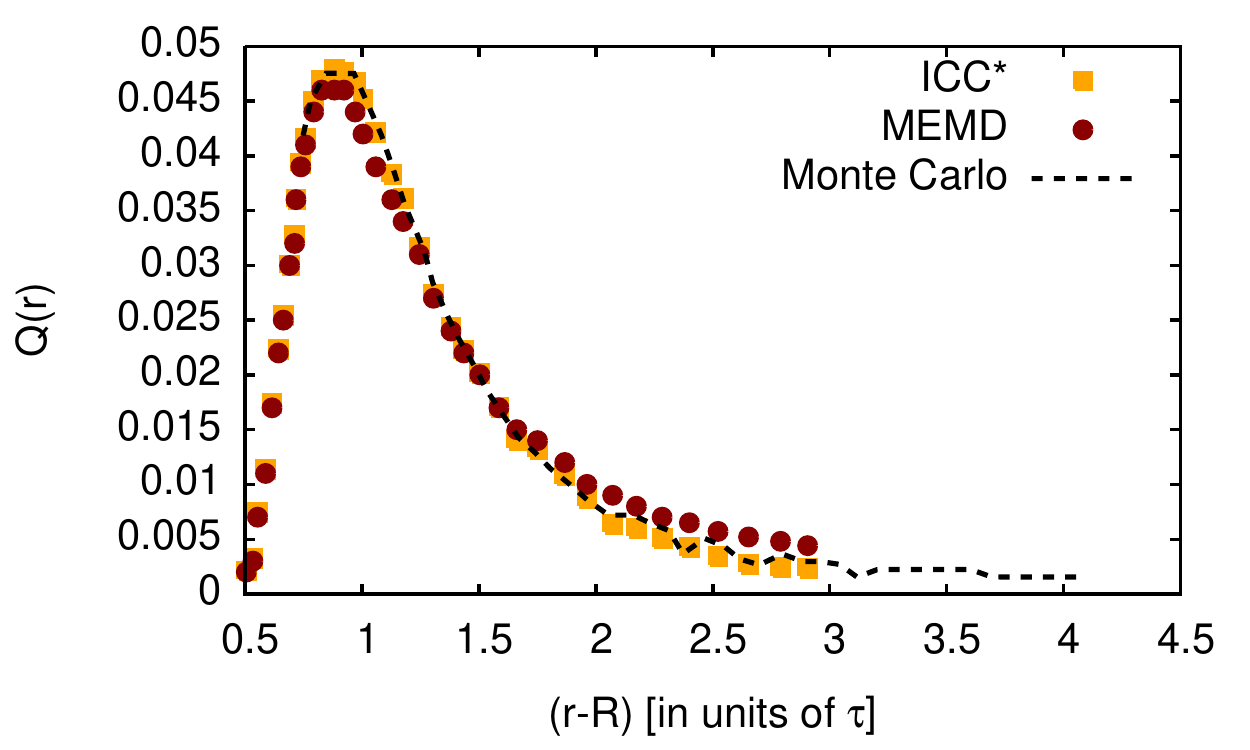}
 \label{fig:dynamic-sphere-data}
}
 \caption{(a) A dielectric sphere of $\varepsilon_C=2$ is placed in the center of the box and salt ions assemble in bulk water $\varepsilon_W=80$ around it. (b) The ion density as a function of the radial distance is analyzed and compared to simulations using the \icc{} algorithm and a Monte Carlo simulation.}
\label{fig:dynamic-sphere}
\end{figure}

Both adapted algorithms (initial and thermodynamics) match the established methods and theoretical predictions to the expected precision. As long as the systematic limitations (see~\ref{sec:limitations}) are kept in mind, the algorithm produces valid results.

\section{Performance and limitations}
\label{sec:performance}
\label{sec:limitations}

\begin{figure}
 \centering
 \includegraphics[width=\plotwidth]{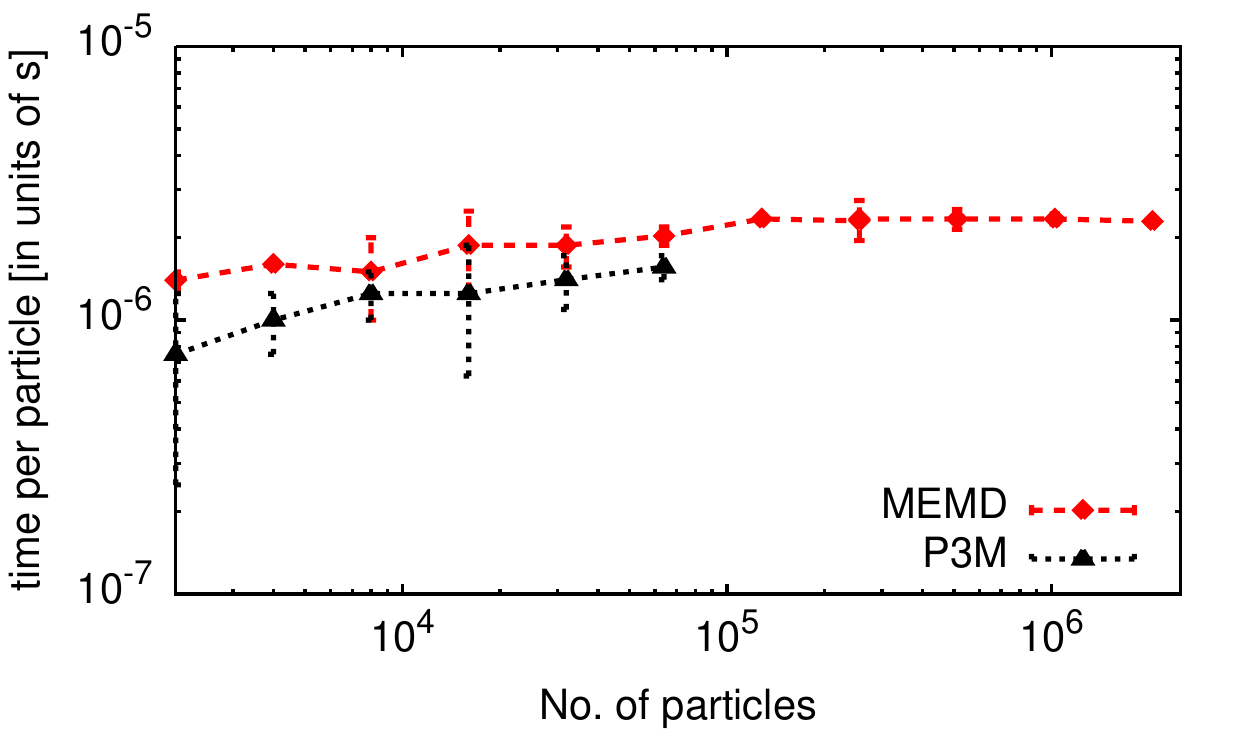}
 \caption{For constant density, the algorithm scales linearly with the particle number from $10^3$ to $2\cdot10^6$ particles. For this system, the implementation is slower than the optimized \pppm{} solver, but the inclusion of dielectric changes comes at no extra cost.}
 \label{fig:scaling}
\end{figure}

The scaling with the number of particles follows the linear $\mathcal{O}(N)$ theoretical prediction as can be seen in figure~\ref{fig:scaling}. The strong parallel scaling works but shows some communication overhead. This is due to the large amount of data being exchanged ($\Vect{B}$-fields, $\Vect{D}$-fields, electric currents and charges for all lattice sites at domain boundaries).

In comparison to other methods, \memd{} performs acceptably. At a comparable RMS force error of $10^{-3}$, the tuned \pppm{} implementation in \es{} runs at about $1.5$ times as fast (see figure~\ref{fig:scaling}). All scalings and simulations were done within the \es{} package~\cite{arnold13a}. Considering that \pppm{} does not include dielectric interfaces or variations and has been highly optimized, this is an acceptable loss in speed. A more thorough performance analysis and comparison to other methods is presented within the library for Scalable Fast Coulomb Solvers, \scafacos{}~\cite{misc-scafacos,arnold13b}.

The \memd{} algorithm comes with some systematic limitations. First, as pointed out in section~\ref{sec:periodicity}, the system should not have an externally driven dipole moment, i.e.\ an electric current. Second, while not impossible, highly inhomogenous systems will reduce the precision noticeably and the speed drastically (see section~\ref{sec:error}). Third, fixed charges within the system (i.e.\ charges that are forced to stay at a predefined position) should be avoided since we noticed that the system equilibrates more slowly and these charges give rise to systematic errors in their close vicinity. Since the algorithm will only propagate fields created from electric currents within the system, and non-moving charges do not produce any current, this will lead to the effect that the influence of a charge is lessened. Pseudo-fixed charges enclosed in a strong potential are a way around this problem. Lastly, for simulations that require a very high precision in force, the linear interpolation scheme is not feasible (see section~\ref{sec:error}).

\section{Conclusion}

We have extended the existing \memd{} algorithm to deal with spatially varying dielectric properties. In the theoretical part, we have analytically proven that the algorithm reproduces correct behavior, conserves energy and reproduces the expected equations of motion. We have introduced an error estimate and a correction for the dipole moment influence arising from periodic boundary conditions. In the simulational part, we have shown that our implementation of the algorithm can reproduce analytical solutions and simulation results based on the induced charge algorithm \icc{}. We have also tested and verified the linear scaling of the implementation.

The extended \memd{} algorithm is an interesting method, and able to dynamically deal with charges in locally varying properties with very little computational overload. It is of high interest to researchers who study electrolytic systems and highly charged objects, such as DNA strands or colloids. The method has already been applied successfully to several research projects, and publications are in preparation.

The implementations in the software packages \es{}~\cite{misc-espressomd} and \scafacos{}~\cite{misc-scafacos} are freely available and parallelized.

\begin{acknowledgments}

We would like to thank Igor Pasichnyk, Stefan Kesselheim, and Anthony Maggs for fruitful discussions.

This work was supported by the DFG through the SFB 716, the Volkswagen foundation, and the German Ministry of Science and Education (BMBF) under grant 01IH08001.

\end{acknowledgments}



\end{document}